\documentclass[aps,prb,amssymb]{revtex4}

\usepackage{graphicx}
\usepackage{color}

\def\ep {\epsilon}

\def\evh {\epsilon_{vH}}
\def\e2 {\epsilon-\epsilon_k}
\def\be {\begin{equation}}
\def\ee {\end{equation}}
\def\bea {\begin{eqnarray}}
\def\eea {\end{eqnarray}}
\def\om {\omega}

\def\cd {c^{\dagger}}

\def\si {\sigma}

\begin{document}

\centerline{ Published in Europhysics Letters
(EPL), 144, 36001 (2023) ; DOI 10.1209/0295-5075/ad0bc6}
\title{Zeeman dependence of the quasiparticle scattering rate and ARPES
in copper oxides and related materials}

\author{George Kastrinakis }

\affiliation{
   Institute of Electronic Structure and Laser (IESL),
Foundation for Research and Technology - Hellas (FORTH),
N. Plastira 100, Iraklio, Crete 70013, Greece$^*$ }

\date{3 November 2023}

\begin{abstract}
Within a strongly interacting Fermi liquid framework, we calculate
the effects of the Zeeman energy $\omega_H$ for a finite magnetic field,
in a metallic system with a van Hove peak in the density of states,
located close to and below the Fermi surface. We find that the chemical
potential increases with the square of $\omega_H$. We obtain a characteristic
quasiparticle scattering rate linear in the maximum of $\omega_H$ and 
temperature, both in the normal and
the d-wave superconducting state. We predict that ARPES experiments in copper
oxides, and related compounds, should
be able to elucidate this behavior of the scattering rate, and in particular,
the difference between spin up and down electrons.

\end{abstract}

\maketitle


{\bf Introduction.}
- The quasi-particle scattering rate (SR) is very charactheristic of the 
very nature of the electronic system in question \cite{abri}, and connected 
with its spectral properties. 
It is also inextricably linked to the 
resistivity and the optical conductivity. In refs. \cite{gk0,gk1} in 1997, 
for a pure Fermi liquid (FL), with
a van Hove singularity, located at the energy $\ep_{vH}$ below the chemical 
potential $\mu$, we had predicted a SR linear
in the maximum of $T$ and $\ep$. Namely for $T>\{x_0/4,\ep\}$, and for
$\ep>\{x_0,T\}$, with $x_0=\mu-\ep_{vH}$. 
Angle-resolved photoemission spectroscopy (ARPES) data \cite{lu} showed that 
$x_0$ is in the range 5-30 meV (and even smaller) for several cuprates. 
The $\ep$-linear 
behavior is limited above by the system bandwidth, while the $T$-linear one
is limited above by the melting of the crystal, or by some other phase
transition which may alter the value of $x_0$.
This SR yields directly a $T$-linear resistivity, and a $1/\ep$ 
optical conductivity, for the appropriate $T,\ep$ ranges \cite{gk0,gk1}.
E.g. the scale of $x_0$ is fully consistent with the onset of $T$-linear 
resistivity.

ARPES experiments have the capacity to yield
the quasi-particle SR, as a function of temperature $T$
and energy $\ep$, for momenta anywhere close to the Fermi surface (FS).
High resolution
ARPES experiments on the cuprates, which demonstrated the $T,\ep$-linear
SR, started with Valla et al. \cite{valla} in 1999 for the normal
state. For the d-wave superconducting state, the linearity of the SR
was pioneered in refs. \cite{john,kami,yusof}.

Recently, two different ARPES experiments
on cuprates, and related materials, in finite magnetic field $H$, have
appeared \cite{ryu,huang}. Finite $H$ is expected to introduce a broadening 
of the SR
lineshape. Of course, this experimental novelty is still at an initial
stage, and its full potential is to be seen.

Herein we further develop our
FL model for quasiparticle scattering \cite{gk0,gk1,gk2}.
The extension of the model here for finite $H$ comprises
only the Zeeman energy $\om_H$, and omits any dependence on orbital effects, 
and on other order parameters, such as density waves of any kind.
We provide concrete expressions for $\om_H <1$, which
can be tested against ARPES data. The possible challenges of extracting 
the SR are to be resolved within the experimental procedure. 

In the following, we first 
discuss the model, the chemical potential and, briefly, the 
density of states (DOS) for finite $H$. With these at hand, we proceed 
to discuss the quasiparticle SR, and related predictions for
ARPES experiments. Finally, we conclude.

{\bf Model and chemical potential for finite $\om_H$.}
- We wish to model the copper oxides. We consider the single band
Hamiltonian with a {\em generic} and non-separable
potential $V_{q}$ 
\bea
H = \sum_{k,\si} ( \ep_{k} - \mu ) \;  \; \cd_{k,\si} c_{k,\si}
+ \frac{1}{2} \sum_{k,p,q,\si,\si'} V_{q} \;
\cd_{k+q,\si} \cd_{p-q,\si'} c_{p,\si'}c_{k,\si}  \;\;  . \;\;
\eea
$\cd_{k,\si}$ creates an electron of momentum $k$ and spin $\si$, $\mu$
is the chemical potential, and we use the relevant band structure
\be
\ep_k = -2t( \cos k_x+\cos k_y )-4 t' \cos k_x \cos k_y
-2t'' \{ \cos (2k_x)+\cos (2k_y) \}   \;\; ,\;\; k_x,k_y=-\pi \rightarrow \pi
\;\; ,\;\; \label{disp}
\ee
which yields both the correct FS, and a van Hove singularity
in the DOS, located at $\ep=\evh<\mu$, corresponding to
$k=(\pm \pi,0),(0,\pm \pi)$. The electron-electron interaction results
in pinning $\evh$ slightly {\em below} $\mu$ for
a broad range of dopings, which may extend fron the underdoped to the
overdoped regime in various cuprates \cite{gk1}. Early experimental 
evidence from ARPES can be found in \cite{lu}.

We wish to calculate $\mu_H=\mu(\om_H)$ for finite $H$.
For this, we need a self-consistent calculation of the ground state energy,
taking into account that the spin-dependent dispersion becomes
\be
\ep_{k,\si} = \ep_k \mp \om_H \;\; , \;\; \si=\uparrow,\downarrow \;\; , \;\;
\ee
and $\ep_k$ given by eq. (\ref{disp}). This results in the FS expanding
and shrinking for spin $\si=\uparrow,\downarrow$, respectively, such that the
total filling factor remains constant with $\om_H > 0$.

We perform a calculation which is simpler than the full Eliashberg-type
one. We work in the static limit (without any frequency dependence).
We use the Hartree-Fock approximation, which allows to calculate
the ground state energy at temperature $T=0$ as \cite{hcom}
\be
E  = \sum_{k,\si} (\ep_{k,\si}-\mu_H) \; n_{k,\si}
- \frac{1}{2} \sum_{k,q,\si} V_{q} \; n_{k+q,\si} \;n_{k,\si}
+\frac{1}{2} \sum_{k,p,\si,\si'} V_{q=0} \; n_{k,\si} \;n_{p,\si'}
\;\; . \;\;
\ee
Superconducting correlations
are omitted. The occupation factors $n_{k,\si}$ take only the values 0 and 1
\cite{zcom}.
The total filling factor is $n=\sum_{k,\si} n_{k,\si} $.
We use the potential
\be
V_q=\sum_{i=1}^{4}
\frac{V_0}{a_0^2 +  \xi^2 (\vec{q}-\vec{Q}_{i})^2 } \;\;. \label{pote}
\ee
The peaks of $V_q$ are symetrically located at $Q_i= (\pm \pi,\pm \pi)$.
We typically take $\xi=2$, $a_0=1$, $V_0=6 t$. $V_q$ may be considered as an
effective screened potential, e.g. like the potentials in the full
self-consistent solutions of the Eliashberg equations - c.f. ref. \cite{gk1}
and therein, and ref. \cite{scal}. This is a strong coupling approach, with
full momentum dependence. We use a $N \times N$, $N=240$, discretization of
the 2-d Brillouin zone. The filling $n$ is held fixed in the calculation,
which yields the {\em self-consistent evaluation} of $n_{k,\si}$,
$\mu_H$ and the respective $\si=\uparrow,\downarrow$ FS's.
The latter are shown in Figs. 1,2, and they are determined from the condition
\be
\Xi_{k,\si}= 0  \;\; , \;\;
\ee
where $\Xi_{k,\si}$ is the renormalized dispersion
\bea
\Xi_{k,\si}= \frac{ \partial E}{\partial n_{k,\si}}
= \ep_{k,\si}-\mu_H+ \Sigma_{1\si}(k,0)  \;\; , \;\;  \\
\Sigma_{1\si}(k,0) = \sum_q \; V_q \; n_{k+q,\si} - V_{q=0} \; n  \;\; . \;\;
\eea
The condition $\Xi_{k,\si}>0$ yields $n_{k,\si}=0$. The (real) self-energy
$\Sigma_{1\si}(k,0)$ here is assumed to correspond to the $\ep=0$ limit
of the real part of the fully $\ep$-dependent self energy
$\Sigma_{1\si}(k,\ep)$ appearing in eq. (\ref{enw}) and further on.

\begin{figure}[tb]
\begin{center}      
  \centering

  \includegraphics[width=6.0truecm,angle=-90]{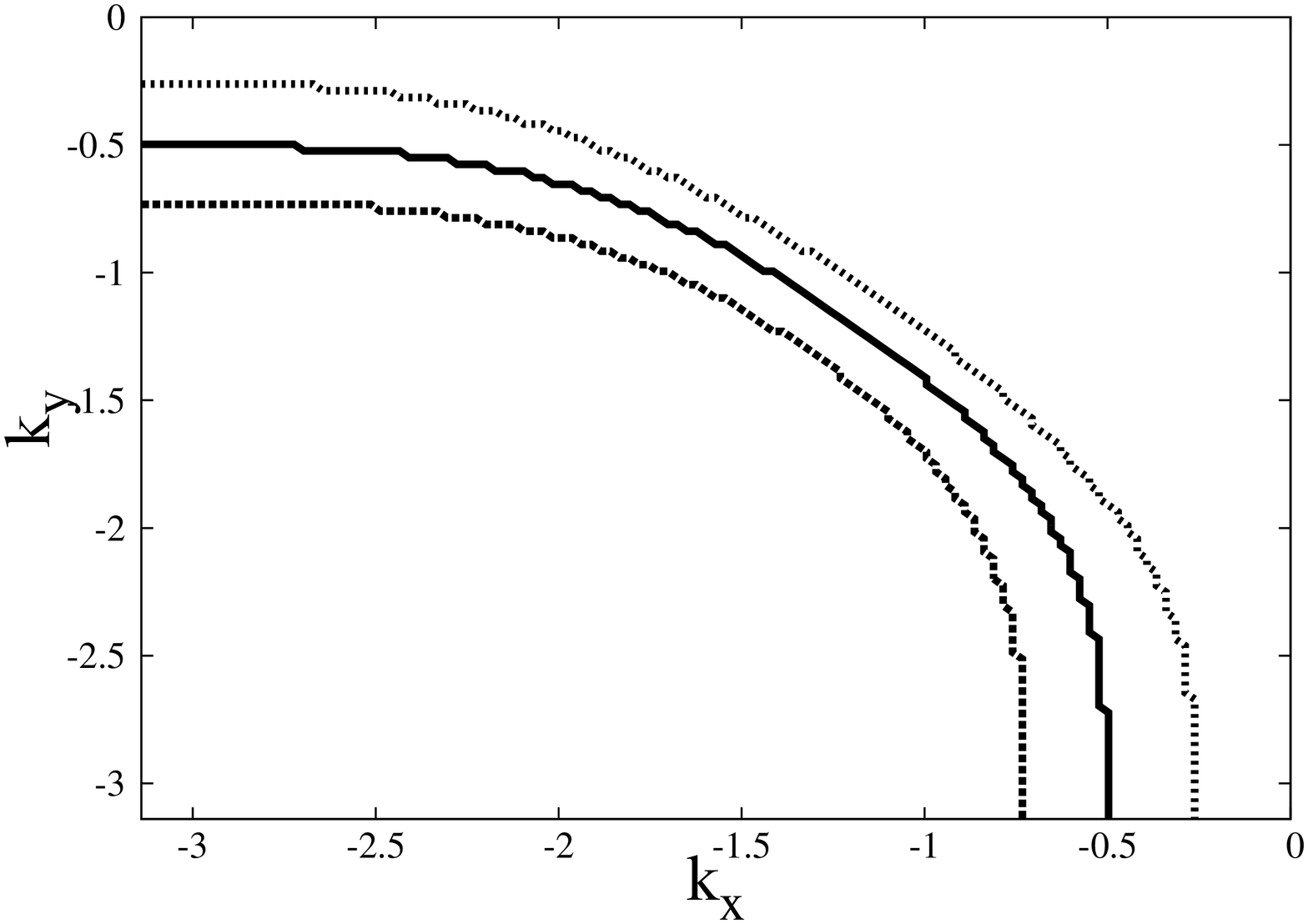}

\caption{ Fermi surface (FS) for $t'=-0.2t$, $t''=0.15t$ and
  $V_q$ of eq. (\ref{pote}), and filling $n=0.81$. The upper right corner
  is the point $\Gamma$ (0,0), and the lower left corner is the point $M$
  $(-\pi,-\pi)$. The middle line corresponds to $\om_H=0$.
Lower line $\om_H=0.4 \; t$ spin-up, upper line $\om_H=0.4 \; t$ spin-down.
Also c.f. text. }

  \includegraphics[width=6.0truecm,angle=-90]{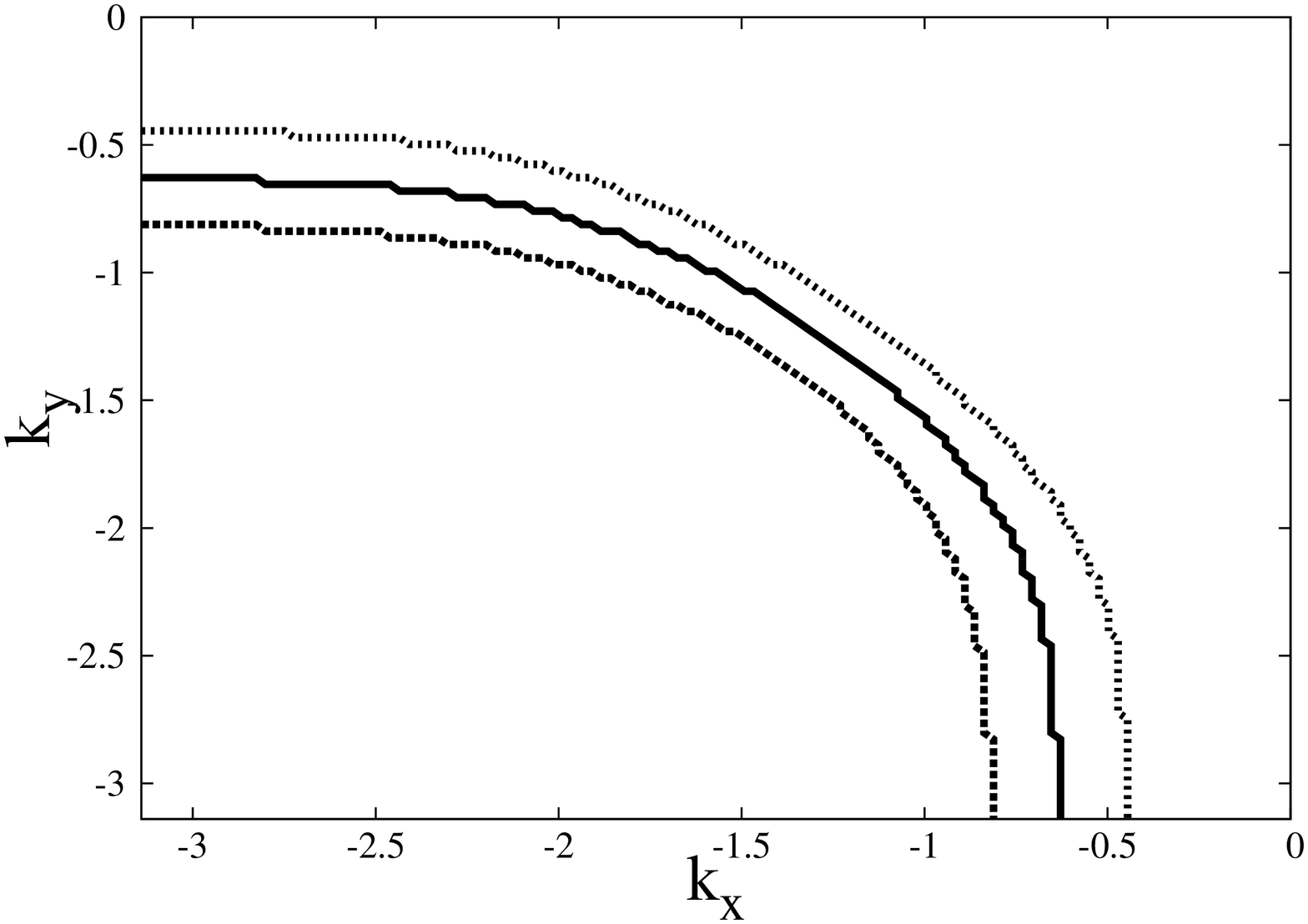}

\caption{ Fermi surface for the same parameters (and geometry) as above, 
and filling $n=0.91$. Middle line $\om_H=0$.
Lower line $\om_H=0.4 \; t$ spin-up, upper line $\om_H=0.4 \; t$ spin-down.
Also c.f. text. }

\end{center}
\end{figure}

As can be seen in Fig. 3, we obtain a parabolic dependence
of $\mu_H$ with $\om_H$
\be
\mu_H = \mu_0 + a \; \om_H^2   \;\; , \;\;  \om_H < \om_*(n) \;\; , \;\;
\label{muhh}
\ee
where the filling dependent $\om_*(n)$ is the value beyond which
the FS of the spin down electrons
shrinks below the points X $(\pm \pi,0)$ and Y $(0,\pm \pi)$. The constant
$a$ depends on the filling factor (and other band parameters), as shown in
Table 1. Though the data shown is for a particular set of band parameters,
eq. (\ref{muhh}) holds in general (with modified $\om_*$'s etc.), as we have
seen in our numerical calculations. We also consider the case
$\mu_H-\mu_0 = a \; \om_H^2 < \om_H/4$. This is true for $\om_H < \om_0$
with
\be
\om_0= \frac{1}{4 a} \;\;. \;\;  \label{muh2}
\ee
In this case ($a \; \om_H^2$) can be ignored compared to $\om_H $, and this is
relevant for the subsequent discussion of the SR. Further
\be
\om_*(n) = \om_0(n)-\delta \om(n) \;\; , \;\; \delta \om(n) \ll \om_*(n)
 \;\;. \;\;  \label{muh3}
 \ee
Values of $\om_0$ and $\delta \om$ also appear in Table 1.

\begin{table}\centering

\begin{tabular}{|c|c|c|c|c|c|} \hline
  
$ \text{filling } $n$ $ & $\mu_0$ $(t)$ & $a$ $(t^{-1})$ & $ \om_*$ $(t)$
& $ \om_0$ $(t)$ & $\delta\om = \om_0-\om_*$ $(t)$ \\ 

\hline
$ 0.71 $  &  -1.1806  &  0.53053 & $ \sim 0.4 $  & 0.471  & 0.071  \\

$ 0.76 $  &  -1.0727  &  0.40521 &  $ \sim 0.55 $ & 0.617  & 0.067  \\

$ 0.81 $  & $ -0.9528 $ & $ 0.32191  $ & $ \sim  0.75 $ & $ 0.777 $ & 0.027 \\

$ 0.835 $ & $ -0.8868 $ & $ 0.28429  $ & $ \sim 0.85 $ & $ 0.879 $ & 0.029 \\

$ 0.86 $ & $ -0.8195 $ & $ 0.26061  $ & $  \sim 0.95 $ & $ 0.959 $ & 0.009 \\

$ 0.885 $ & $ -0.7476 $ & $ 0.23405  $ & $ \sim 1.05 $ & $ 1.07 $ & 0.02 \\

$ 0.91 $ & $ -0.6736 $ & $ 0.21385 $ & $\sim 1.15 $ & $ 1.17 $ & 0.02 \\

\hline
\hline

\hline

\end{tabular}

{ Table 1. Parameters related to $\mu_H$, as in eqs. (\ref{muhh}),
(\ref{muh2}), (\ref{muh3}),
for $t'=-0.2t$, $t''=0.15t$ and $V_q$ of eq. (\ref{pote}), for filling
factors $n=0.71 - 0.91$. Also c.f. Figs. 1, 2 and text. } 
\end{table}

\begin{figure}[tb]
\begin{center}      
  \centering

  \includegraphics[width=7truecm,angle=-90]{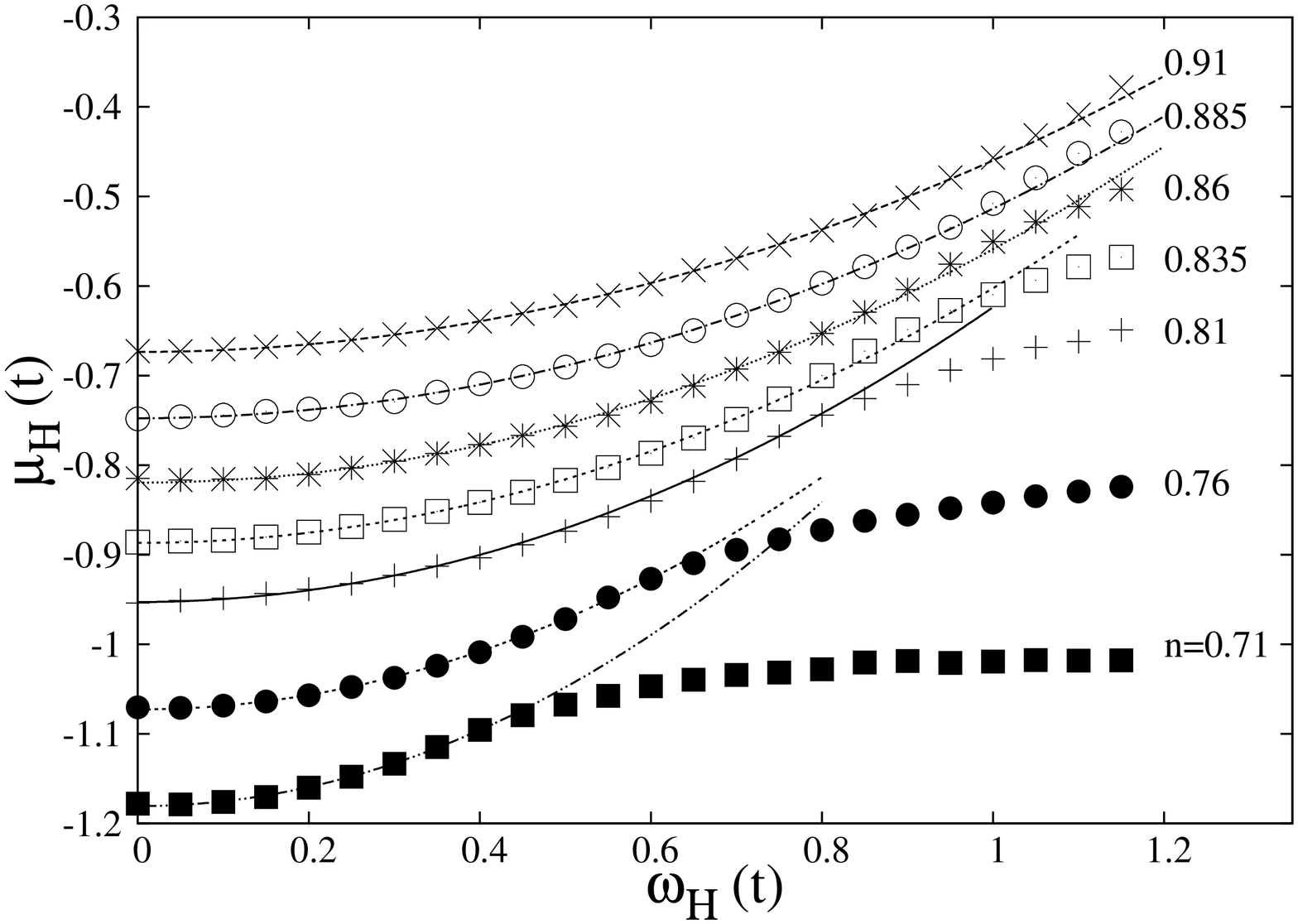}    

\caption{  Chemical potential $\mu_H$ $(t)$ vs. 
$\om_H$ $(t)$ for $t'=-0.2t$, $t''=0.15t$ and 
$V_q$ of eq. (\ref{pote}). The curves bottom to top correspond to the filling
factors $n=0.71,0.76,0.81,0.835,0.86,0.885,0.91$. The symbols come out of the
self-consistent numerical calculations. The lines are fits of eq. (\ref{muhh})
with the parameters given in Table 1. The 
$\om_H^2$ dependence is valid up to $\om_*=0.4 t$ for $n=0.71$. 
$\om_*$ steadily increases for increasing $n$.  
Also c.f. text. }

\end{center}
\end{figure}

As shown e.g. in Fig. 4 of ref. \cite{gk1}, the DOS $N(\ep)$ can be
approximated as a smooth background with a single peak for $\om_H=0$, located
close to and below the FS. This is known both from numerous ARPES
experiments \cite{lu} on various materials,
and from several calculations (like in \cite{gk1} and therein). For finite
$\om_H > 0$ the DOS splits into two distinct $\uparrow,\downarrow$ components,
self adjusting in order to conserve the total spectral weight.

The  $\om_H^2$ dependence of $\mu_H$ is also obtained through an analytical
calculation, wherein the DOS $N(\ep)$ is taken as a constant background, with
an additional high peak slightly below the FS for $\om_H=0$. Then we require
conservation of the spectral weight 2 $\int_{-A}^{\mu_0} d\ep \; N(\ep)=
\int_{-A}^{\mu_H}  d\ep \; \{ N(\ep+\om_H) + N(\ep-\om_H) \}$.
Here $(-A)$ is the lower energy cut-off.

{\bf Quasiparticle scattering rate.}
- We calculate the quasiparticle SR within a FL
framework, as in our previous work \cite{gk0,gk1,gk2}.
Our approach is also within the framework of conserving Baym-Kadanoff
approximations \cite{bk}. The derivation for the SR is
{\em valid both in the normal and the d-wave superconducting state}.
In \cite{gk2} we analysed relevant limitations for a s-wave superconducting
state. Further, we assume that any $T>0$ dependence of $\mu_H$ is
{\em subleading} compared to the other effects of $T>0$.

The SR is a genuine FL type one, in the sense that for
$x$=max$\{T,\ep\} \rightarrow 0$, it is proportional to $x^2$ as in
\cite{gk0,gk1,gk2}.
As a result, there is no concomitant analyticity issue (as e.g. in
Krammers-Kronig relations connecting the real and imaginary parts of the
self-energy).
Only within appropriate $T$ and $\ep$ bounds is the rate linear in $x$, and
possibly linear in $\om_H$ (as shown below).

We start from the Eliashberg equations, as e.g. written in \cite{gk2},
and follow the same notation.
The quasiparticle self-energy is $\Sigma(k,\ep) = \text{Tr} \; V \; G$, where
$V(q,\om)$ is the effective potential and $G(k,\ep)$ the Green's function.
We use the expression for the SR, i.e. the imaginary part
of $\Sigma(k,\ep)$ in eq. (18) therein
\be
\Sigma_2(k,\ep) \simeq \frac{1}{2}\sum_q V_q^{(1)}(0) B(k-q,\ep-w_{k-q,\ep})
\; w_{k-q,\ep} \; A(w_{k-q,\ep},\ep,T) \;\;.\;\;
\ee
Here $V_q^{(1)}(0)=\partial \; \text{Im} V(q,\om=0) / \partial \om $,
and we have the thermal factor
\be
A(w_{k,\ep},\ep,T) = \coth \left( \frac{w_{k,\ep}}{2T} \right)
+ \tanh \left( \frac{\ep-w_{k,\ep}}{2T} \right) \;\;, \;\;   \label{tfa}
\ee
which satisfies
\be
A(w_{k,\ep},\ep,T) = 0 \;\;, \;\; \ep < w_{k,\ep} < 0  \;\;.\;\;
\ee 
The energy factor $w_{k,\ep}$ is given by
\be
w_{k,\ep}=\frac{\ep+\mu-\ep_k-\Sigma_1(k,\ep)-D_1(k,\ep)}
{1-\partial_{\ep} \Sigma_1(k,\ep) -\partial_{\ep} D_1(k,\ep) } \;\;.\;\;
\label{enw}
\ee
$\ep$ is the quasiparticle energy,
$\Sigma_1(k,\ep)$ is the real part of $\Sigma(k,\ep)$, and $D_1(k,\ep)$
the real part of the superconducting gap. For a small gap, we have the
conditions $D_i^2(k,\ep) \ll f_i^2(k,\ep)$ (c.f. between eqs. (10) and (11)
in \cite{gk2} for $f_i$ and $D_i$, $i=1,2$). Then we get
\be
B(k,\ep)=1-\frac{D_1(k,\ep) \; f_1(k,\ep) + D_2(k,\ep) \; f_2(k,\ep) }
{ f_1^2(k,\ep) + f_2^2(k,\ep) } \;\; .\;\;
\ee
In the normal state $B(k,\ep)=1$.

For {\em finite} $H$ the above $k$-dependent quantities and $\mu$ become
$\om_H$-dependent. {\em Omitting the $\om_H$-dependence} of 
$D_1(k,\ep)$ we now have

\be
w_{k,\ep,H}^{\si}=\frac{\ep+\mu_H-\ep_k \pm \; \om_H -\Sigma_{1\si}(k,\ep)
  -D_1(k,\ep)}
{1-\partial_{\ep} \Sigma_{1\si}(k,\ep) -\partial_{\ep} D_1(k,\ep) } \;\;.\;\;
\ee

From our numerical calculations we see that  $\Sigma_{1\si}(k,0)$ depends
on $\om_H$, namely the part $S_{\si}(k) = \sum_q \; V_q \; n_{k+q,\si}$.
It can be very well approximated by
\be
S_{\si}(k) = S_{0}(k) + g_{\si}(k) \; \om_H  \;\; , \;\;
\om_H < \om_f(n)   \;\; , \;\; \label{sdep}
\ee
with the filling dependent $\om_f(n) < \om_*$, as shown in fig. 4.
In table 2 we show the FS averaged values
$S_0=<S_0(k)>_{FS}$, $g_{\si}=<g_{\si}(k)>_{FS}$ of the parameters of eq.
(\ref{sdep}).

We draw the attention of the reader to the analysis of the relevant
inequalities (19)-(24) in \cite{gk2}.
Repeating the analysis therein, for $T>\ep,w_{k,\ep,H}^{\si}$ the
SR is linear in $T$
\be
\Sigma_{2\si}^{R}(k,\ep,H) \simeq -T \; \sum_q V_q^{(1)}(0) \;
B(k-q,\ep-w_{k-q,\ep,H}^{\si}) \;\; .\;\;  \label{tsig}
\ee

\begin{table}\centering

\begin{tabular}{|c|c|c|c|c|c|} \hline
  
 \text{filling} $n$ & $S_0$ $(t)$ &  $g_{\uparrow}$ &  $g_{\downarrow}$ &
 $\om_f$ $(t)$ \\
\hline

 0.71  & 0.1867  & 0.1840  &  -0.1755 &  0.4   \\

 0.76  & 0.2088  & 0.1628  &  -0.1420 &  0.45   \\  

 0.81  & 0.2259  &  0.1639  &  -0.1379 &  0.75   \\  

 0.835 & 0.2358 &  0.1615  &  -0.1355 &   0.8  \\  
 
 0.86  &  0.2463 &  0.1580 &  -0.1324 &   0.8  \\  

 0.885 &  0.2572 &  0.1539 &  -0.1287 &  0.9   \\ 
 
 0.91  &  0.2689 &  0.1485 &  -0.1247 &  1.0   \\  

\hline
\hline

\hline

\end{tabular}

{ Table 2. Parameters related to $S_{\si}$, averaged over the FS, 
as in eq. (\ref{sdep}),
for $t'=-0.2t$, $t''=0.15t$ and $V_q$ of eq. (\ref{pote}), for filling
factors $n=0.71 - 0.91$. Also c.f. text. } 
\end{table}

\begin{figure}[tb]

\begin{center}      
  \centering

 \includegraphics[width=7truecm,angle=-90]{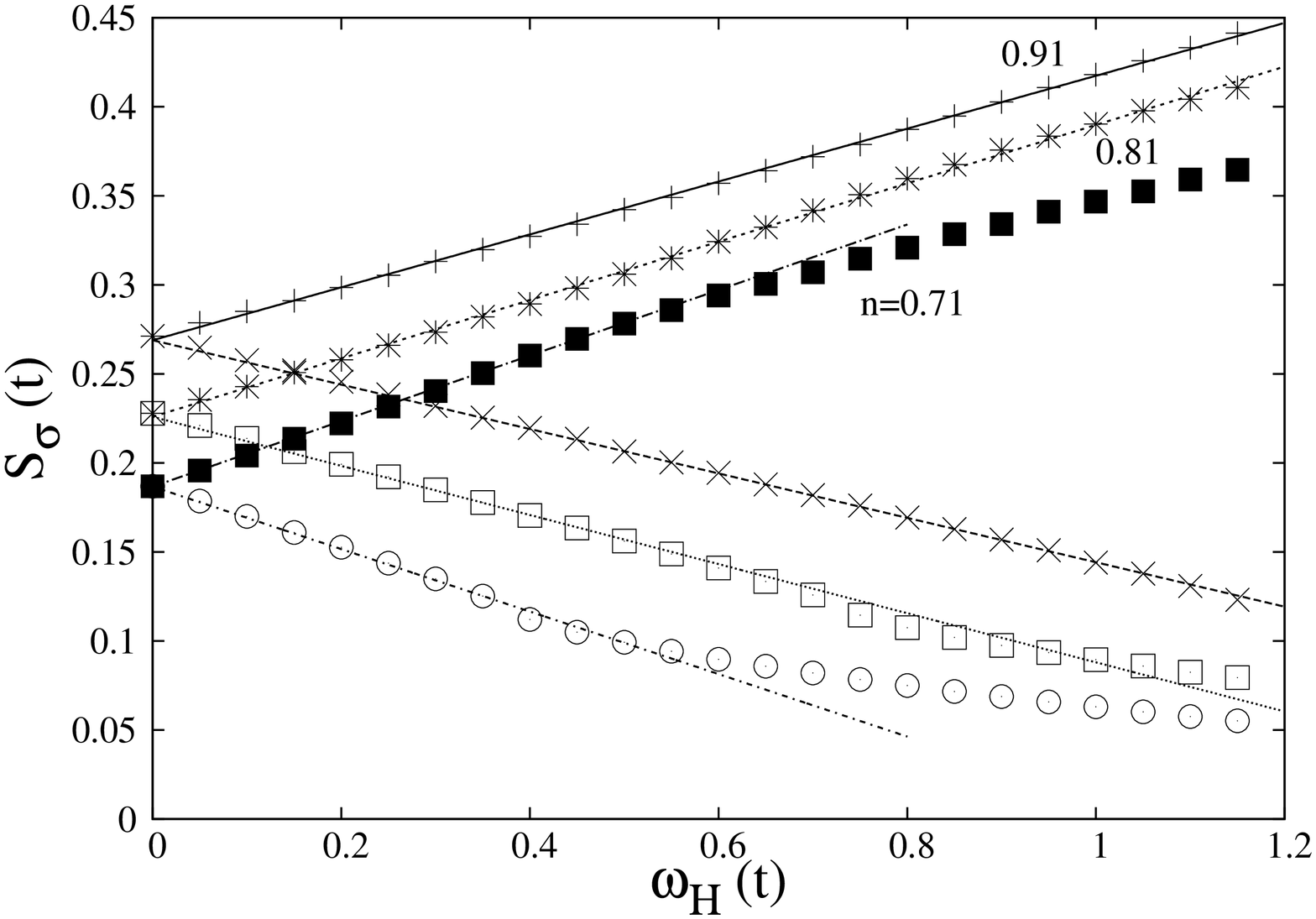}
  
\caption{ Real part of the self-energy $S_\si$ $(t)$ vs. $\om_H$ $(t)$ 
in eq. (\ref{sdep}), averaged over the FS, for $t'=-0.2t$, $t''=0.15t$ and 
$V_q$ of eq. (\ref{pote}). The curves bottom to top correspond to the filling
factors $n=0.71,0.81,0.91$. The symbols come out of the self-consistent 
numerical calculations. The straight lines are fits of eq. (\ref{sdep})
with the parameters given in Table 2, showing that the 
$\om_H$-linearity is valid at least up to $\om_H=0.5 t$ here. 
For $n=0.71$ the linear fit fails
for smaller $\om_H$ than the other 2 cases.  $S_\si$ enters in the
expression for the scattering rate in eq. (\ref{sel2}), thus contributing
to its $\om_H$-linear dependence. Also c.f. text. }

\end{center}
\end{figure}

In the limit $\om_H > T $ we obtain
\be
\Sigma_{2\si}^R(k,\ep,H) \simeq -\frac{ d_{k}^{\si}}{2}
 \sum_q V_q^{(1)}(0) \; B(k-q,\ep-w_{k-q,\ep,H}^{\si}) \; w_{k-q,\ep,H}^{\si} \;\;. \;\;
\ee
Here $d_{k}^{\si}$=1+$<$tanh[$(\ep-w_{k-q,\ep,H}^{\si})/(2T)$]$>$ is a factor
between 1 and 2. Also note that
\be
\partial_\ep \Sigma_{1\si}(k,\ep) = -b_k < 0 \;\;,\;\; \ep < \ep_o=O(t) \;\;.\;\;
\ee

Then we obtain 
\be
\Sigma_{2\si}^R(k,\ep,H) \simeq -\frac{ d_{k}^{\si}}{2}
\sum_q V_q^{(1)}(0) \; B(k-q,\ep-w_{k-q,\ep,H}^{\si}) \;
\frac{ \ep+\mu_H-\ep_{k-q} \pm \; \om_H -\Sigma_{1\si}(k-q,\ep)}
{1+b_{k-q}}   \;\;.\;\;   \label{sel2}
\ee
There is interference between $\mu_H$, the pure Zeeman term
$\pm \; \om_H$, and the term $g_{\si} \; \om_H$. It should be possible
to probe the rates
$\Sigma_{2\si}^R(k,\ep,H)$ by performing ARPES experiments in adequately
strong field $H$. In particular, it should also be possible to measure
the pure Zeeman dependence, with the assumption that $d_{k}^{\si}=d_{k}$
has a {\em negligible dependence on $\om_H$}. The appropriate quantity is
the difference
\be
\Sigma_{2 \uparrow}^R (k,\ep,H) - \Sigma_{2 \downarrow}^R (k,\ep,H)
= -d_{k} \; \om_H \; \sum_q \frac{ V_q^{(1)}(0) }
{1+b_{k-q}} \; B(k-q,\ep-w_{k-q,\ep}) \; l_{k-q} \;\;,\;\;
\ee
with
\be
l_k= 1+ \frac{1}{2} \{ g_{\uparrow}(k) - g_{\downarrow}(k) \}  \;\;.\;\;
\ee
In particular, in the normal state this formula becomes
\be
\Sigma_{2 \uparrow}^R (k,\ep,H) - \Sigma_{2 \downarrow}^R (k,\ep,H)
= -y_{k} \; \om_H \; \; ,\;\;\;
\ee
with
\be
y_{k} = d_{k} \; \sum_q \frac{ V_q^{(1)}(0) }
{1+b_{k-q}} \; l_{k-q} \;\;.\;\;   \label{siga}
\ee
This is to be compared with the total prefactor for the case $T>\ep,\om_H$
in $\Sigma_2^{R}(k,\ep)=-x_{k} \; T$, c.f. eq. (\ref{tsig}),
\be
x_{k} = \sum_q V_q^{(1)}(0) \;\;.\;\;  \label{sigb}
\ee
Here the two prefactors $y_{k}$ and $x_{k}$ are similar in magnitude.
This should be visible in ARPES experiments.

Further, we examine the quantity
\be
s_{\si}(k,H) = \sum_q V_q^{(1)}(0) \; \Xi_{k-q,\si} \;\;,\;\;  \label{lsi}
\ee
which satisfies $\Sigma_{2 \si}^R (k,\ep=0,H) \propto s_{\si}(k,H)$.
Assuming that the $\om$-dependent $V(q,\om)$ (compatible with $V_q$ in
eq. (\ref{pote}) ) is
\be
V(q,\om) = \sum_{i=1}^{4}
\frac{V_0}{a_0^2 +  \xi^2 (\vec{q}-\vec{Q}_{i})^2 - i \; \om/\om_F } \;\;, \;\;
\ee
we have
\be
V_q^{(1)}(0) =\frac{V_0}{\om_F} \; \sum_{i=1}^{4}
\frac{1}{ \big[ a_0^2 +  \xi^2 (\vec{q}-\vec{Q}_{i})^2 \big]^2 } \;\;. \;\;
\label{pote2}
\ee 
The small fermionic energy is typically $\om_F=10-40$ meV \cite{gk1}.
Here we take $\om_F=t/10$. For e.g. $t=250$ meV, this yields $\om_F=25$ meV.

The numerical calculations yield
\be
s_{\si}(k,H) = s_{0}(k) + f_{\si}(k) \; \om_H  \;\; , \;\;  \om_H < \om_1
\;\; , \;\;   \label{labs}
\ee
typically with $\om_1 \leq \om_*(n)$, in accordance with eq. (\ref{sdep}).
The FS averaged coefficients $s_\si$ and $f_\si$ are shown in Table 3.
Note that $f_{\uparrow}<0$ and $f_{\downarrow}>0$.

We note that there is an {\em asymmetry} between up and
down spin electrons for finite $\om_H$, as is
evident from figures 4 and 5, and from tables 2 and 3. This asymmetry is
due to the {\em non-linear dispersion $\ep_k$}. It is more pronounced for
$s_{\si}(k,H)$ rather than for $S_{\si}(k)$. 

\begin{table}\centering

\begin{tabular}{|c|c|c|c|} \hline
  
 \text{filling} $n$ & $s_0$ &  $f_{\uparrow}$ &  $f_{\downarrow}$  \\
\hline

 0.71  &  1.8021 &  -1.8039  &  1.1347  \\

 0.76  &  1.6147 &  -1.7839  &  1.2399  \\

 0.81  &  1.4095 &  -1.7604   &  1.3055  \\

 0.86  &  1.1713 &  -1.7107  &  1.3734  \\

 0.91  &  0.9224  & -1.7086  & 1.4195  \\

\hline
\hline

\hline

\end{tabular}

{ Table 3. Parameters related to $s_{\si}$, averaged over the FS, as 
in eq. (\ref{lsi}),
for $t'=-0.2t$, $t''=0.15t$ and $V_q^{(1)}(0)$ of eq. (\ref{pote2}), with 
$\om_F=t/10$ and for filling
factors $n=0.71 - 0.91$. $f_{\uparrow}<0$ and $f_{\downarrow}>0$. Also c.f. text. } 
\end{table}

\begin{figure}[tb]
\begin{center}      
  \centering

  \includegraphics[width=7truecm,angle=-90]{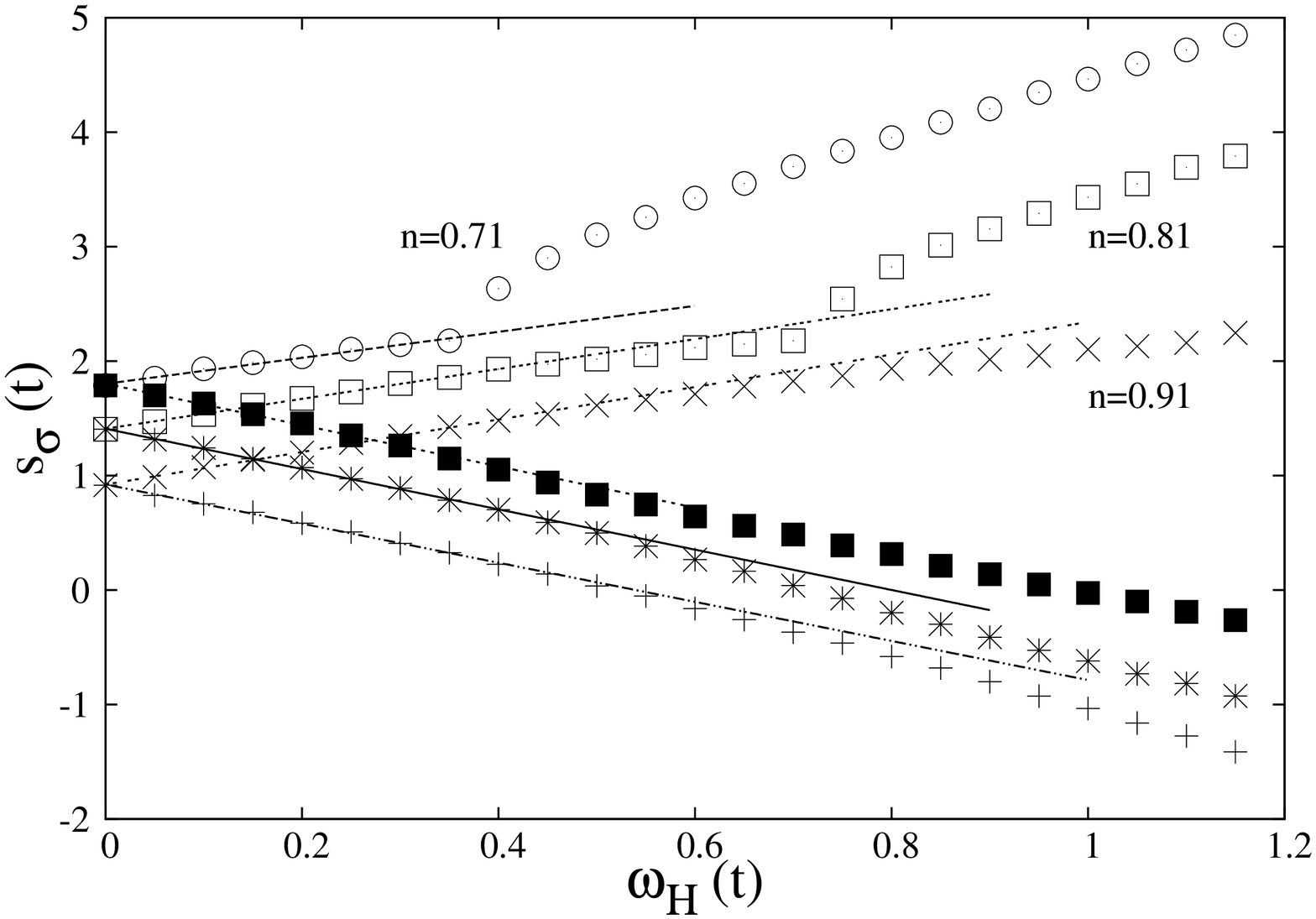}

\caption{  $s_{\si}$ $(t)$ (an estimate of the scattering rate) vs. 
$\om_H$ $(t)$ in eq. (\ref{lsi}), averaged 
over the FS, for $t'=-0.2t$, $t''=0.15t$ and $V_q^{(1)}(0)$ 
of eq. (\ref{pote2}). The curves top to bottom correspond to the filling
factors $n=0.71,0.81,0.91$. The symbols come out of the self-consistent 
numerical calculations. The straight lines are fits of eq. (\ref{labs})
with the parameters given in Table 3, showing that the 
$\om_H$-linearity is valid at least up to $\om_H=0.35 t$ here. 
For $n=0.71$ the linear fit fails
for smaller $\om_H$ than the other 2 cases. Also c.f. text. }

\end{center}
\end{figure}

{\bf Overview. }
- Based on our earlier work, we calculate the SR 
in a finite magnetic field, taking only the Zeeman energy $\om_H$
dependence into account, within a purely Fermi liquid framework. The DOS has
a strong van Hove peak (slightly below the FS).

We make specific predictions for the quasi-particle SR, which 
can be probed by ARPES experiments. We find a characteristic
linear in max $\{T,\om_H\}$ dependence. 

The related compounds BaFe$_2$As$_2$ \cite{bafe} and CeCoIn$_5$ \cite{ceco}
have shown a combination of $T$-linear resistivity, a van Hove singularity
located close to the FS, and a $T$-linear SR (modified
possibly by structural phase transitions).
Our results should apply in these, and other related, compounds as well.

\vspace{.1cm}
$^*$ e-mail : kast@iesl.forth.gr ; giwkast@gmail.com


\begin{thebibliography}{99}
\bibitem{hcom} The momentum sums are normalised by the number of total
momentum grid points $N^2$ - c.f. below.
\bibitem{zcom} As a result of $n_{k,\si}=0$ or 1 {\em only}, it is impossible
to have a quasiparticle weight $z_k<1$ around the FS. However, since we
are {\em not} concerned about such fine detail here, this is acceptable.
\end{thebibliography}
\end{document}